\documentclass[aps,twocolumn,prd,nofootinbib,letterpaper]{revtex4}
\usepackage{graphicx}
\usepackage{amssymb}
\usepackage{epstopdf}
\usepackage{amsmath}
\usepackage{subfigure}
\usepackage{dcolumn}
\usepackage{bm}
\usepackage{latexsym}
\usepackage{epsf}
\usepackage{graphics}
\usepackage{amsfonts}
\usepackage[mathscr]{eucal}
\usepackage{enumerate}
\usepackage{hhline}
\usepackage{array}
\usepackage{tabularx}

\newcommand{\beq}{\begin{equation}}
\newcommand{\eeq}{\end{equation}}
\newcommand{\beqa}{\begin{eqnarray}}
\newcommand{\eeqa}{\end{eqnarray}}

\newcommand{\gs}{g_{\rm s}}
\newcommand{\alphap}{\alpha'}
\newcommand{\Mpl}{M_{\rm Pl}}
\newcommand{\dd}{{\rm d}}
\newcommand{\coupl}{g_{\phi\chi}}

\newcommand{\uone}{u_{1}}
\newcommand{\ucrit}{u_{1,{\rm crit}}}
\newcommand{\ueps}{u_{1,\epsilon}}

\newcommand{\phiin}{\phi_{\rm in}}
\newcommand{\phihit}{\phi_{\rm hit}}

\newcommand{\phieps}{\phi_{\epsilon}}
\newcommand{\phicrit}{\phi_{\rm crit}}
\newcommand{\phisr}{\phi_{\rm sr}}

\newcommand{\ie}{\emph{i.e.}{}}

\newcommand{\Trh}{T_{\rm rh}}
\newcommand{\ktot}{k_{\rm tot}}
\newcommand{\ksr}{k_{\rm sr}}

\begin{document}

\title{Reheating in a Brane Monodromy Inflation Model}

\author{Robert H. Brandenberger$^{1)}$, Anke Knauf$^{1)}$, 
Larissa C. Lorenz$^{2,1)}$}

\affiliation{\qquad $^1$~ Department of Physics, McGill University,
Montr\'eal, QC, H3A 2T8, Canada \\
$^2$~Institut d'Astrophysique de Paris, 98bis boulevard Arago, 75014 Paris,
France.}

\date{version September 16, 2008}

\begin{abstract}

We study reheating in a recently proposed brane ``monodromy inflation''
model in which the inflaton is the position of a $D4$ brane on a 
``twisted torus''. Specifically, we study the repeated collisions between 
the $D4$ brane and a $D6$ brane (on which the Standard Model fields 
are assumed to be localized) at a fixed position
along the monodromy direction 
as the $D4$ brane rolls down its potential. We find that there is no
trapping of the rolling $D4$ brane until it reaches the bottom of its
potential, and that reheating 
is entirely described by the last brane encounter. Previous collisions 
have negligible effect on the brane velocity and hence on the reheat 
temperature. In the context of our setup, reheating is efficient and
the reheat temperature is therefore high.
\end{abstract}

\maketitle

\section{Introduction}
\label{sec:introduction}

Among the string theory constructions aiming for plausible early Universe 
scenarios, brane inflation models are a popular and promising subclass
(for reviews see e.g. \cite{Braneinflation}). 
They often identify the scalar field $\phi$ responsible for the early 
inflationary expansion with the position of a $D$-brane of suitable 
dimensionality [or possibly the distance between several (anti-)$D$-branes] 
in the extra, compactified dimensions. This, however, leads to a 
geometrical upper bound on the field range accessible to such a stringy 
inflaton: At best, $\phi$ can travel over the entire extension of the 
compactified dimension(s), but in most scenarios only part of this range 
is actually suitable for supporting inflation. Taking into account the inflaton's canonical normalization, its field range was
thought to correspond to ``small field inflation'' (field values
smaller than the Planck mass). Since the Lyth bound \cite{Lyth} directly 
relates the distance travelled in field space to the contribution of
tensor modes (gravitational waves) to the observed fluctuations in the
cosmic microwave background (CMB) temperature maps, brane inflation 
scenarios have therefore been known to predict an unobservably low contribution 
from tensor perturbations \footnote{As discussed in \cite{Martineau},
nonlinear effects due to the primordial scalar metric fluctuations
lead to tensor modes, and a lower bound on this contribution can be
derived.}. 

It turns out that this obstruction to having a large contribution
to CMB anisotropies from tensor modes is far from being a no-go theorem. 
As has recently been emphasized \cite{Eva2}, more complicated backgrounds 
than warped Calabi--Yau manifolds (which abound in brane inflation models) 
make it possible for the inflaton field to traverse a certain geometric 
cycle numerous times, adding up to a large distance in field space.
The model proposed in \cite{Eva2} is based on a ``twisted torus''
background for ten-dimensional type IIA superstring theory and is reviewed
in the following section of this paper. The inflaton field $\phi$
corresponds to the position of a $D4$ brane wrapping the twisted torus
numerous times. 

To obtain a successful inflationary model, it is crucial to consider
the exit from inflation and the energy transfer between the inflaton and
Standard Model matter fields. In Type IIA theories, Standard Model
matter must be localized on branes. In this paper, we study
the reheating process assuming that Standard Model matter is confined
to a $D6$ brane localized at a certain point along the twisted torus.
As the $D4$ brane unwinds, it hits the $D6$ brane numerous times.
After the final intersection (after it has unwound completely), the $D4$ brane 
will come to rest intersecting the D6 in (3+1) spacetime dimensions.
However, there is the danger that
the $D4$ brane might get trapped by the $D6$ at earlier intersections 
due to strings stretching between them. These strings become massive as the branes separate again,
and might thus prevent a graceful exit from inflation. In this paper,
we show that this trapping does not occur. Reheating is dominated
by the final intersection. We estimate the reheat temperature
after inflation and find that it is high.

Reheating in
previously proposed brane inflation models has been studied in a 
large number of papers. Reheating in a brane world model with bulk
inflaton was studied in \cite{Himemoto} (see also \cite{Rouzbeh}). In
a brane-antibrane inflation model the reheating process was investigated
by means of the tachyon condensation process in \cite{CFM} (see also
\cite{Takamizu} for a study of reheating in a non-inflationary
brane-anti brane model). A large
body of work is devoted to studying reheating in two throat
brane inflation models in which the Standard Model lives in a
deeper throat than the one in which the brane-antibrane annihilation
process takes place \cite{BBC,ReBBC}. Reheating through the
relaxation of a throat was proposed in \cite{Frey}. Our mechanism
is based on the trapping mechanism by enhanced symmetry
states analyzed in \cite{Kofman:2004yc,Watson} (see also \cite{Greene} for
a more recent study). Key to our analysis is the production of
open string states in brane interactions, a process studied in
detail in \cite{Leblond}. Reheating in an earlier $D6-D4$ brane 
inflation model was analyzed in \cite{Easson}. A study of
reheating in a $D3-D7$ inflation model was presented in \cite{BDD}.

The outline of this paper is as follows: In the following two sections we
review the brane inflation model of \cite{Eva2}, with particular
emphasis on the expressions for the potential energy function in the
two limiting regions of field space. In Section 4, we derive the
relevant equations valid during the slow-roll phase, before going
on to our main topic, namely  the
interaction between the $D4$ and $D6$ branes during the unwinding
of the former, and show that (at least for the parameter values
preferred in \cite{Eva2}) no trapping occurs. This is done in
Section 5. In the final section we
then estimate the reheat temperature.
 
\section{The Model}

\subsection{The IIA String Background}

The specific background we consider was constructed in 
Ref.~\cite{Silverstein:2007ac} and used for inflationary model 
building in Ref.~\cite{Eva2}. This scenario relies on 
ten-dimensional type IIA string theory with six dimensions compactified 
on a nilmanifold, that is, a ``twisted torus''. This kind of manifold is 
T-dual to type IIB string theory compactified on a torus with 
Neveu-Schwarz (NS) flux. Under T-duality, the NS field becomes part of 
the geometry and leads to a non-trivial fibration of the T-duality cycles 
over the base. To solve the supergravity equations of motion, this 
background has to include Ramond-Ramond (RR) two- and four-form flux 
as well as orientifold planes, but they will not be of concern to us here\footnote{See \cite{Silverstein:2007ac} for details. The setup presented in \cite{Eva2} is laid out in such a way that the orientifold planes do not interfere with the motion of the $D4$ brane.}.

More precisely, the six-dimensional internal manifold is a product of two 
twisted three-tori. Following the notation of \cite{Silverstein:2007ac}, 
we denote the metric of one of these twisted tori 
[with coordinates $(x,\uone,u_{2})$] by
\begin{eqnarray}\label{metric}
\frac{ds^2_{\rm tt}}{\alpha'}  & = & L^2_{u_1}\,du_1^2 + L^2_{u_2}\,du_2^2 + 
L_x^2(dx'+Mu_1\,du_2)^2,\label{eq:nilmanifoldmetric}
\end{eqnarray}
where $x'=x-(M/2)u_1 u_2$, and $M$ is an integer flux quantum number. 
At fixed $u_1$, the metric of Eq.~(\ref{eq:nilmanifoldmetric}) describes a 
torus in the $(x',u_2)$ direction. At $u_1=0$, this is simply a square torus, 
but moving along the $u_1$ direction, the complex structure $\tau$ of this 
torus changes from $\tau \to \tau+M$ as $u_1\to u_1+1$. The manifold is 
compactified by identifying these two tori, in other words, there is a 
non-trivial monodromy as we go around the $u_1$ direction. To be more 
precise, the manifold is compactified by making the identifications
\begin{eqnarray}\nonumber
  (x,u_1,u_2) &\sim& \left(x+1,u_1,u_2\right),\\
  (x,u_1,u_2) &\sim& \left(x-M/2\,u_2,u_1+1,u_2\right),\label{eq:identifications}\\ 
  \nonumber
  (x,u_1,u_2) &\sim& \left(x+M/2\,u_1,u_1,u_2+1\right)\,.
\end{eqnarray}
This means that the coordinates $(x,u_1,u_2)$ are restricted to the 
interval $[0,1]$, but with a slight abuse of notation we will let $u_1$ 
run over the whole real axis to describe multiple revolutions 
along this direction.

This background admits a state which gives accelerated expansion 
because the negative scalar 
curvature term leads to a positive contribution to the potential energy
\footnote{For a large enough uplift, \cite{Silverstein:2007ac} also requires 
supersymmetry breaking KK monopoles (wrapped 5-branes).}. Supersymmetry is 
broken at a high scale, corresponding to the lowest KK scale of the geometry, because the potential responsible for this is dictated by the curvature of the manifold. 
In the parameter range of \cite{Silverstein:2007ac}, we can assume that the 
curvature remains weak, but some of the toroidal fibres become very small. 
We adopt the point of view that all moduli have been stabilized by fluxes 
as well as the potential due to the non-K\"ahler structure of the background. 
This is justified since the inflationary model building in 
Ref.~\cite{Eva2} was carried out in the region where these 
moduli remain fixed and the only dynamical field is the inflaton $\phi$, 
whose interpretation we now discuss.

\subsection{Monodromy inflation}

To study inflation in this background, one can imagine to wrap a 
$D4$ brane [with its (4+1)-dimensional 
worldvolume] along the $u_2$ direction, its remaining (3+1) dimensions 
filling the uncompactified dimensions of spacetime. This is not a 
supersymmetric setup because the background we place the $D4$ in breaks supersymmetry by itself. The $D4$ will aim to minimize its action, and 
therefore its worldvolume. As mentioned above, the area of the $(x', u_2)$ 
torus is minimized for $u_1=0$, so the $D4$ that wraps a one-cycle inside this torus will unwind by traversing the 
$u_1$ cycle several times, until its wrapping number has decreased to 1 and 
it reaches the position $u_1=0$. Using a suitable renormalization derived 
from the requirement of a canonic kinetic term for the inflaton, the brane 
position in the $u_1$ direction is now interpreted as the inflaton field 
$\phi$. Since the nilmanifold has a non-trivial monodromy under 
$u_1\to u_1+1$ [see Eq.~(\ref{eq:identifications})], $\phi$ does not 
come back onto itself after completing one cycle. It can therefore cross 
a large field range although the compact coordinate $u_1$ is restricted to 
lie in $[0,1]$\footnote{See the comment following 
Eq.~(\ref{eq:identifications}) for the use of the range of $u_1$.}. 
In Ref.~\cite{Eva2} it was shown that inflation in this 
model takes place for super-Planckian field values with a potential 
$V(\phi)\propto\phi^{2/3}$, and that $N\sim60$ e-folds of expansion can 
be achieved before the slow-roll conditions are violated.

However, to complete this inflationary scenario, one has to specify the 
mechanism of reheating describing the transition to Standard Model cosmology. 
To this end, a few more model building ingredients are necessary. In type II 
theories, the Standard Model is usually constructed on the intersection of 
stacks of branes on a toroidal orbifold (see e.g. \cite{Marchesano} for a review). In type IIA, 
candidate branes [with at least (3+1) worldvolume dimensions] are $D4$, 
$D6$ and $D8$-branes. $D4$ and $D8$-branes are usually disregarded as 
they would have to wrap a one- or five-cycle, respectively, both of which 
are homologically trivial in Calabi-Yau manifolds. The most successful
minimal supersymmetric standard models (MSSM) require several stacks of 
intersecting $D6$ branes, typically using all internal dimensions. We will 
be less ambitious here and consider only a single $D6$ brane that is 
wrapped on the second nilmanifold\footnote{We could also imagine replacing 
this nilmanifold by an ordinary toroidal orientifold without twisted fibres. 
Twisting is only essential on the nilmanifold on which we want to achieve 
monodromy inflation. The motivation in Ref.~\cite{Eva2} to 
have both the nilmanifolds twisted is to make them interchangeable under an orbifold projection and 
possibly use combinations of their respective coordinates [$(x',u_1,u_2)$ 
and $(\tilde{x}',\tilde{u}_1,\tilde{u}_2)$] as candidate inflatons.}, which 
has a metric $d\tilde{s}^{2}_{\rm tt}$ analogous to 
Eq.~(\ref{eq:nilmanifoldmetric}) with a second set of coordinates 
$(\tilde{x}',\tilde{u}_1,\tilde{u}_2)$. Again, this is not a supersymmetric 
setup, but we assume the $D6$ has already minimized its worldvolume, sitting 
at a meta-stable equilibrium position. We can treat the $D6$ as a probe in the 
same way we treat the D4 as a probe in this background geometry. We 
choose our setup such that the $D4$ and the $D6$ overlap in three spatial 
directions but are orthogonal on the internal manifold (this would preserve supersymmetry in flat space). We therefore 
do not expect any tachyons to appear in the open string 
spectrum, even after the $D4$ is wrapped on a twisted fibre. An open string stretching between the two branes would only become tachyonic when the branes come very close to each other. At this point, it is the local geometry that matters and the branes are mutually BPS on scales on which the curvature of the background can be neglected. The existence of this tachyon should therefore be insensitive to the global supersymmetry breaking by the background.

During inflation, the $D4$ unwraps and crosses the $D6$, which we choose 
to localize at $u_1=0$, several times. Each time, open string modes 
between the branes are created. As long as the collision is non-relativistic, 
only unexcited massless strings are produced \cite{McAllister:2004gd}. 
This will indeed turn out to be the case here, the velocity of the inflaton 
along the slow-roll trajectory being extremely small. These inter-brane 
strings, however, acquire a mass when the branes move apart from each other 
again following each collision, and therefore create an attracting force 
between the $D4$ and the $D6$. This attractive force is in competition 
with the force (\ie{} the inflaton's potential) that tries to unwind the 
$D4$, aiming to minimize its worldvolume. Note that this additional potential 
is the main difference to the case studied in Ref.~\cite{Kofman:2004yc}, 
where only the string-induced potential was present.

One might worry that this could lead to a trapping of the $D4$ on the $D6$ 
before a sufficient amount of inflationary expansion has been achieved 
\cite{Kofman:2004yc}. It turns out, however, that the number density 
$n_{\chi}$ of the produced strings at each collision is so small (owing to 
its dependence $n_{\chi}\propto|\dot{\phi}|^{3/2}$, with the $D4$'s kinetic 
energy being extremely small during inflation) that their contribution to 
the effective potential does not stop the inflaton from rolling down towards 
the minimum at $u_1=0$. Due to the monodromy in the $\uone$ direction, new 
strings are produced at each crossing, in principle allowing for an 
accumulative effect. But as long as each $\uone$ turn corresponds to 
several e-folds of slow-roll inflation on the $V(\phi)\propto\phi^{2/3}$ 
potential, the earlier collisions' string modes have already been diluted 
to negligible density when the branes meet again. Only towards the end of 
the unwrapping process does slow-roll break down, and the string densities 
of the last few collisions may add up. We will discuss this effect in detail 
below.

Once the $D4$ has unwound completely, it oscillates around the $D6$, 
reheating it and providing the phenomenological connection to the Universe's 
subsequent evolution. Simultaneously, when $\uone$ becomes small, the notion 
of the renormalized inflaton field $\phi$ in terms of the coordinate $u_1$ 
changes. As a consequence, during this final stage (generically after the 
previous slow-roll phase has ended) the potential is quadratic, 
$V(\phi)\propto\phi^{2}$. To estimate the temperature of reheating in this 
model, we need to determine the velocity of the inflaton when it reaches its 
potential minimum at $u_1=0$ (corresponding to $\phi_{0}=0$) and therefore 
the minimum of the $D4$-brane's worldvolume.

\section{Inflaton normalization and potential}\label{sec:inflaton}

''Monodromy inflation'' is a large field inflation model, \ie{} accelerated 
expansion of the Universe proceeds while the inflaton $\phi$ moves from a 
large initial value (measured in Planck units) towards smaller field values. 
In Ref.~\cite{Eva2} it was shown that, in order to avoid 
destabilizing the moduli, the inflaton field has to start below a certain 
geometry-imposed maximum value $\phi_{\rm max}$. Above this field
value the inflaton energy density is so large that the dynamics of the
other moduli cannot be neglected. We do not discuss this 
restriction in detail here and only note that it translates into 
$\phi_{\rm max}/\Mpl\simeq\mathscr{O}(10)$ or less, corresponding roughly 
to the same order of magnitude of turns $\ktot$ in the $\uone$ direction.

The action for the inflaton is derived from the Dirac-Born-Infeld (DBI) world
sheet action of the $D4$ brane in the presence of the non-trivial background
geometry
\beqa \label{action1}
S_{D4} \, &=& \, \int \frac{d^5 \zeta}{(2 \pi)^4 (\alpha')^{5/2}} e^{- \Phi} \\
& & \times \sqrt{det(G_{MN} + B_{MN})\partial_{\alpha} X^M \partial_{\beta} X^N} \, ,
\nonumber
\eeqa
where $X^M(\zeta_{\alpha})$ are the embedding coordinates of the brane, the
world sheet coordinates being denoted by $\zeta_{\alpha}$ (Greek indices
are world sheet coordinates, capital Latin indices are bulk spacetime 
coordinates). The bulk metric and bulk NS two-form are $G_{MN}$ and $B_{MN}$,
respectively, so that the argument of the square root gives the pullback of these fields onto the brane worldvolume. 
The dilaton is denoted by $\Phi$, and the string scale is
determined by $\alpha'$ (employing the standard notation from string theory).

Taking the brane to be extended in our three spatial dimensions, uniform in
the $u_2$ direction, and located at the position $u_1(y)$ in the monodromy
direction (the coordinates $y$ being our four-dimensional spacetime 
coordinates), the above action reduces to \cite{Eva2}
\beqa \label{action2}
S_{D4} \, &=& \,  \frac{1}{(2 \pi)^4 g_s (\alpha')^2} \int d^4 y \sqrt{-g_4} \\
& & \times \sqrt{ (\beta L_u^2 + L_x^2 M^2 (u_1)^2) 
\left( 1 - \alpha' \frac{(L_u)^2}{\beta} {\dot u_1}^2 \right)} \, , \nonumber
\eeqa
where $\beta = L_{u_2}/L_{u_1} \equiv L_u^2/(L_{u_1})^2$. $L_{x},L_{u_{1}}$ and $L_{u_{2}}$ denote the size of the twisted torus in the respective directions, hence $\beta$ measures the ''anisotropy'' between the $u_{1}$ and $u_{2}$ directions and $L_{u}$ an average over the two. $g_s$ is the string coupling constant whose value is set by the expectation value of the dilaton
$\Phi$, $g_4$ is the 
determinant of the induced spacetime metric, and the overdot
indicates the derivative with respect to physical time.

As is apparent from (\ref{action2}), the field $u_1(y)$ is not canonically
normalized. For applications to cosmology we need to transform to the
corresponding canonically normalized field $\phi(u_1)$, in terms of which
the action will then be given by
\beq
S_{D4} \, = \, \int d^4 y \,\sqrt{-g_4} \,
\left[ \frac{1}{2}\, {\dot \phi}^2 - V(\phi) \right] \, . 
\eeq
Expanding the action (\ref{action2}) up to two derivatives, we obtain
the following form for the potential
\beq
V(\phi) \, = \, \frac{\beta^{1/2} L_u}{(2 \pi)^4 g_s \alpha'^2}\,
\sqrt{1 + \frac{M^2 L_x^2}{\beta L_u^2}\, u_1^2(\phi)} \, .
\eeq

In the small and large field regions, the conversion between the original field $u_1$ and the canonically normalized field $\phi$ is of an explicit and simple form. We now list the results (from \cite{Eva2}) and
the corresponding potentials.

\subsection{At small field values}

In the regime $\uone<\ucrit$, where $\ucrit$ (see Ref.~\cite{Eva2}) 
is given by
\beq\label{eq:ucrit}
\ucrit\,\sim\,\frac{\sqrt{\beta}}{M}\left(\frac{L}{L_{x}}\right)^{3/2},\qquad L^{3}=L_{u}^{2}L_{x}\,,
\eeq
the potential takes the form
\beq\label{eq:potentialsmall}
V^{\phi<\phicrit}(\phi) \, = \, \frac{m^{2}}{2}\,\phi^{2}\,,
\eeq
where the ''mass'' $m$ is given in terms of the background parameters by
\beq\label{eq:mscale}
m^{2} \, = \, \frac{M^{2}}{\alpha'}\frac{L_{x}^{4}}{L^{6}}\,.
\eeq

In the small field regime, the relation between $\uone$ and $\phi$ is 
linear and given by \cite{Eva2}
\beq\label{eq:defphismall}
\frac{\phi}{\Mpl} \, = \, 
\frac{(2\pi)^{3/2}\gs^{1/2}L_{u}^{3/2}}{\beta^{1/4}L^{3}}\,\uone\,.
\eeq
The values of $\phi$ corresponding to $\uone<\ucrit$ are (for the
parameter values chosen in \cite{Eva2} which yield a successful
inflationary model) much smaller than the four-dimensional Planck mass
$\Mpl$, and hence do not lie in the slow-roll regime for inflation.  

\subsection{At large field values}

In the region $\uone\gg\ucrit$, the potential takes the form
\beq\label{eq:potential}
V(\phi) \, = \, \mu^{10/3}\phi^{2/3}\,,
\eeq
where the mass scale $\mu$ is given in terms of the background parameters
by
\beq\label{eq:mu}
\left(\frac{\mu}{\Mpl}\right)^{10/3}=\left(\frac{3}{2}\right)^{2/3}\frac{1}{(2\pi)^{8/3}}\left(\frac{M^{2}\beta}{\alpha'^{5}\Mpl^{10}\gs^{2}}\right)^{1/3}\frac{L_{x}}{L}\,.
\eeq
If we now calculate the first two slow-roll parameters for the potential of 
Eq.~(\ref{eq:potential}), we obtain
\beqa\label{eq:srparameters}
\epsilon \, &\equiv& \, \frac{\Mpl^{2}}{2}\left(\frac{V_{\phi}}{V}\right)^{2}
\, = \, \frac{2}{9}\frac{\Mpl^{2}}{\phi^{2}}\,, \nonumber \\
\eta \, &\equiv& \, \Mpl^{2}\left(\frac{V_{\phi\phi}}{V}\right)
\, = \, -\epsilon \, . 
\eeqa
Hence, it follows that the slow-roll conditions
$\epsilon,|\eta| \, < \, 1$ both hold until the field reaches the value $\phieps$
given by
\beq\label{eq:phieps}
\frac{\phieps}{\Mpl} \, = \, \sqrt{\frac{2}{9}}\,.
\eeq
For the parameter values used in \cite{Eva2}
(see Appendix A), the breakdown of the slow-roll
approximation occurs in the large field range, i.e.
\beq
\phicrit \, < \, \phieps \, ,
\eeq
and we have $\phi \, > \, \phieps$ for most of the region $\phi\gg\phicrit$.
Hence, slow-roll inflation can occur on the potential (\ref{eq:potential}).

For completeness, let us also mention the relation between $\phi$ and $\uone$ in
the large field regime \cite{Eva2}:
\beq\label{eq:defphi}
\phi \, = \, \frac{M^{1/2}}{6\pi^{2}}
\frac{L_{u}L_{x}^{1/2}}{\left(\gs\alphap\beta\right)^{1/2}}\, \uone^{3/2} \, .
\eeq

\section{Slow-roll regime}

The effective four-dimensional action for the monodromic inflaton 
$\phi$ is
\beq
\mathcal{S} \, = \, -\int\dd^{4}y\,\sqrt{-g}
\left[\frac{R}{2\kappa}-\frac{g^{\mu\nu}}{2}\,\partial_{\mu}\phi\,\partial_{\nu}\phi+V(\phi)\right],
\eeq
where $\kappa \, = \, 1/\Mpl^{2}$, $R$ is the 4d scalar curvature, and the 
potential $V(\phi)$ is given by Eq.~(\ref{eq:potential}).
Assuming the standard spatially flat FLRW metric, the cosmological evolution 
is then governed by the Friedmann and Klein-Gordon equations,
\begin{eqnarray}
H^{2}&=&\frac{1}{3\Mpl^2}\left[\frac{\dot{\phi}^{2}}{2}+V(\phi)\right],\\
-V_{\phi}&=&\ddot{\phi}+3H\dot{\phi}\,.
\end{eqnarray}
Assuming that the slow-roll conditions Eqs.~(\ref{eq:srparameters}) hold, 
these equations are well approximated by
\begin{eqnarray}
H^{2} \, &\simeq& \, \frac{1}{3\Mpl^2}\,V(\phi)\,,\label{eq:Fsr}\\
-V_{\phi} \, &\simeq& \, 3H\dot{\phi}\,.\label{eq:KGsr}
\end{eqnarray}
It follows immediately that the Hubble parameter scales with $\mu$ and $\phi$ as
\beq\label{eq:Hubble}
\frac{H}{\Mpl} \, \simeq \, \frac{1}{\sqrt{3}}
\left(\frac{\mu}{\Mpl}\right)^{5/3}\left(\frac{\phi}{\Mpl}\right)^{1/3} \, .
\eeq
Likewise, combining Eqs.~(\ref{eq:Fsr}) and (\ref{eq:KGsr}) gives for the 
inflaton's velocity
\beq\label{eq:phidotphi}
\frac{\dot{\phi}}{\Mpl^{2}} \, \simeq \, -\frac{2}{3^{3/2}}
\left(\frac{\mu}{\Mpl}\right)^{5/3}\left(\frac{\Mpl}{\phi}\right)^{2/3}.
\eeq
The velocity is negative because the field rolls down the potential towards 
smaller $\phi$ values ($\uone$ decreases from its maximum value $u_{1,{\rm in}}$
to $u_1=0$ as the brane unwraps). The kinetic energy at a given field position 
then is
\beq\label{eq:Ekin}
\frac{E_{\rm kin}(\phi)}{\Mpl^{4}} \, = \, \frac{2}{3^{3}}
\left(\frac{\mu}{\Mpl}\right)^{10/3}\left(\frac{\Mpl}{\phi}\right)^{4/3},
\eeq
while at the same time, the potential energy amounts to
\beq\label{eq:Epot}
\frac{E_{\rm pot}(\phi)}{\Mpl^{4}} \, = \, 
\left(\frac{\mu}{\Mpl}\right)^{10/3}\left(\frac{\phi}{\Mpl}\right)^{2/3}.
\eeq
Comparing Eqs.~(\ref{eq:Epot}) and (\ref{eq:Ekin}), we see that both energy 
contributions scale as $\mu^{10/3}$, but $E_{\rm{kin}}\propto\phi^{-4/3}$ 
and $E_{\rm{pot}}\propto\phi^{2/3}$. Therefore, roughly speaking, 
$E_{\rm pot}>E_{\rm kin}$ for super-Planckian field values. In particular, 
note that $E_{\rm pot}\approx E_{\rm kin}$ (which occurs when 
$\epsilon\simeq1$) around $\phi/\Mpl\simeq\mathscr{O}(1)$, in agreement with 
the value of $\phieps$ from Eq.~(\ref{eq:phieps}).

Re-writing Eq.~(\ref{eq:KGsr}) in terms of the number $N=\int H\dd t$ 
of e-foldings of inflation instead of in terms of cosmic time $t$ allows 
us to integrate and find $N(\phi)$ at a given field value:
\beq\label{eq:Nofphi}
N(\phi) \, = \, \frac{3}{4\Mpl^{2}}\left(\phiin^{2}-\phi^2\right),
\eeq
where we assume that inflation starts at the field value $\phiin$ (thus 
setting the constant $N_{\rm in} \, \equiv \, N(\phiin) \, = \, 0$). 
Inverting Eq.~(\ref{eq:Nofphi}), we find for $\phi$ along the slow-roll 
trajectory
\beq\label{eq:phiofN}
\phi(N) \, = \, \sqrt{\phiin^{2}-\frac{4\Mpl^{2}}{3}\,N}\,.
\eeq
This equation describes the field evolution until we reach $\phieps$. 
Provided that $\phieps \, > \, \phicrit$, the slow-roll phase is followed 
by a period when the potential is still given by (\ref{eq:potential}),
but the motion is too fast to sustain inflation. Then, when the inflaton 
reaches $\phicrit$, the potential becomes quadratic as given by 
Eq.~(\ref{eq:potentialsmall}).

\section{Brane collision}

We now consider the following situation: The $D4$ brane starts out at some 
initial field value $\phiin$ located on the slow-roll trajectory. During
its motion down the potential it will collide several times with the
$D6$ brane. During the collision process, open strings connecting the
two branes will be produced. They will create a restoring force which
opposes the further motion of the $D4$ brane. Below we study the strength of this opposing force. We show that,
at least for the parameter values assumed in \cite{Eva2}, the force
is too weak to trap the $D4$ brane.

Let us follow the motion of the $D4$ brane around the torus in the $\uone$ 
direction and focus on the collision with the $D6$ which occurs at some 
position $\phi=\phihit$. With the slow-roll trajectory still valid, the 
impact velocity $\dot{\phi}_{\rm hit}$ is small, and therefore a 
non-relativistic treatment of the collision suffices and only unexcited strings are produced
\cite{McAllister:2004gd}. In the string theory picture, 
at the moment of collision when the branes coincide, open strings are 
produced with one end attached to each brane. As the $D4$ moves away 
from the $D6$ again, these strings become massive and try to pull the $D4$ 
back towards $\phihit$.

\subsection{Effective field theory description}\label{subsec:fieldtheory}

At the field theory level, one can model a (non-relativistic) collision 
\cite{Kofman:2004yc} by the Lagrangian
\begin{eqnarray} \label{lag}
\mathcal{L}&=&\frac{1}{2}\,\partial_{\mu}\phi\,\partial^{\mu}\phi+V(\phi)\nonumber\\
&&+\frac{1}{2}\,\partial_{\mu}\chi\,\partial^{\mu}\chi-\frac{g_{\phi\chi}^{2}}{2}\left(\phihit-\phi\right)^{2}\chi^{2}\,,
\end{eqnarray}
which describes the coupling of the inflaton $\phi$ to a field $\chi$ that 
becomes massless at the collision point. In the phenomenological picture at hand, 
$\chi$ stands for the lowest energy modes of the string connecting the two
branes. 

The Lagrangian (\ref{lag}) is of the same type which describes the
reheating at the end of inflation in field theory models of inflation.
As was discussed in that context \cite{TB,KLS1,STB,KLS2}, the
equations of motion which follow from (\ref{lag}) for the $\chi$ field
have instabilities which lead to $\chi$ particle production with a
particle number exponentially increasing in time.
Specifically, particle production is concentrated
in time intervals during which the evolution of $\chi$ is non-adiabatic
(see \cite{KLS2} for an in-depth discussion). As shown in
\cite{Kofman:2004yc} and \cite{Watson} this can lead to the stabilization
of moduli fields (like the field $\phi$ in our case) at enhanced
symmetry points. One application of this mechanism is to moduli stabilization
in string gas cosmology \cite{BV,NBV} (see \cite{Subodh} for a 
discussion of this application).
 
Returning to our discussion, the mass of the $\chi$ particles 
increases in proportion to
the distance between the two branes, i.e.
\beq
m^{2}_{\chi}(t) \, = \, g^{2}_{\phi\chi}\left(\phihit-\phi\right)^{2} \, ,
\eeq
where $g^{2}_{\phi\chi}$ is related to the string coupling constant. Thus,
the time-dependent frequency for the $k$th mode of the $\chi$ field
is given by
\beq
\omega(t) \, = \, \sqrt{k^{2}+m_{\chi}^{2}(t)}\,,
\eeq
and one can calculate the ``adiabaticity parameter''
\beq
\left|\frac{\dot{\omega}}{\omega^{2}}\right| \, \approx \, 
\frac{\left|\dot{\phi}\right|}{g_{\phi\chi}}\frac{1}{\left(\phihit-\phi\right)^{2}}\,.
\eeq
This is greater than $\mathcal{O}(1)$ in the interval
\beq\label{eq:deltaphi}
|\Delta\phi| \, \leq \, \sqrt{\frac{|\dot{\phi}|}{g_{\phi\chi}}}\,,
\eeq
where $\dot{\phi}$ is evaluated at the collision time. Taking this
velocity from Eq.~(\ref{eq:phidotphi}) (valid while $\phi$ is 
on the slow-roll trajectory) gives
\beq\label{eq:deltaphi2}
\left|\frac{\Delta\phi}{\Mpl}\right| \, \leq \, 
\frac{\sqrt{2}}{3^{3/4}\coupl^{1/2}}\left(\frac{\mu}{\Mpl}\right)^{5/6}
\left(\frac{\Mpl}{\phi}\right)^{1/3} \, .
\eeq
For the parameter values of \cite{Eva2}, the value of this expression is
much smaller than $1$. It is within this small field range that 
particle/string creation occurs. In particular, we expect this process 
to be instantaneous compared to the Hubble time $H^{-1}$. At the speed 
(\ref{eq:phidotphi}), the field needs a time interval
\beq\label{eq:comparetH}
\Delta t\, H \, = \, \frac{3^{1/4}}{\sqrt{2\coupl}}
\left(\frac{\mu}{\Mpl}\right)^{5/6}\left(\frac{\phi}{\Mpl}\right)^{2/3}
\eeq
(where we have made use of Eq.~(\ref{eq:Hubble}) to determine the Hubble 
parameter) to pass through the field range (\ref{eq:deltaphi}).
This is small compared to one (and hence the time $\Delta t$ short compared to 
the Hubble scale) if the field value $\phi$ satisfies
\beq
\frac{\phi}{\Mpl} \, \ll \, 
\left(\frac{\Mpl}{\mu}\right)^{5/4}\left(\frac{2\coupl}{\sqrt{3}}\right)^{3/4}.
\eeq
For the parameter values discussed in Appendix A this is always the case for 
$\phi$ values sufficiently small such that moduli stabilization is not jeopardized
(see Ref.~\cite{Eva2}). 
Therefore it is justified to treat the process of string creation as 
instantaneous while the inflaton is on its slow-roll trajectory.

The force on the $D4$ brane created by the strings yields a contribution
to the effective potential which determines the motion of the $D4$ brane
 after its encounter with the $D6$ brane. Note, however, that unlike the example 
of Ref.~\cite{Kofman:2004yc}, our Lagrangian (\ref{lag}) already contains a 
bare potential for $\phi$ [given by Eq.~(\ref{eq:potential})] before the 
collision. The total effective potential $V_{\rm eff}(\phi)$ after the brane 
encounter will therefore comprise the old $V(\phi)$ as well as the induced 
string contribution. 

According to Ref.~\cite{Kofman:2004yc}, the number 
density of produced $\chi$ particles in a head-on brane collision is given by
\beq\label{eq:nchi}
n_{\chi} \, = \, 
\frac{\left(\coupl|\dot{\phi}_{\rm hit}|\right)^{3/2}}{\left(2\pi\right)^{3}}\,,
\eeq
where $\dot{\phi}_{\rm hit}$ is the field velocity upon impact. As the $D4$ 
moves past the $D6$, these created particles produce a contribution
to the potential for the $\phi$ field of the form
\beq\label{eq:rhochi}
\rho_{\chi}(\phi) \, = \, \coupl n_{\chi}\left(\phihit-\phi\right).
\eeq
Assuming that we are on the slow-roll trajectory and that therefore 
(\ref{eq:phidotphi}) can be used, this potential becomes
\beq
\rho_{\chi}(\phi) \, = \, \frac{\coupl^{5/2}\Mpl^{4}}{2^{3/2}\pi^3 3^{9/4}}
\left(\frac{\mu}{\Mpl}\right)^{5/2}\left(\frac{\Mpl}{\phihit}\right)
\left(\frac{\phihit-\phi}{\Mpl}\right).
\eeq

It is possible that this new potential energy contribution creates a local minimum in which the inflaton might get trapped.
Our next goal is to evaluate
whether the $D4$ comes to a halt after its encounter with the $D6$ brane
located at $\phihit$.

\subsection{Trapping Effects}

\emph{Before} the $D4$ hits the $D6$, $\phi$ is moving along the slow-roll 
trajectory (\ref{eq:phiofN}) that was obtained from 
Eqs.(\ref{eq:Fsr})-(\ref{eq:KGsr}) with the potential (\ref{eq:potential}). 
In this way, we can calculate the impact velocity $\dot{\phi}_{\rm hit}$ and 
therefore the string production rate at the time of the collision:
\beq
\frac{n_{\chi}^{(1)}}{\Mpl^{3}} \, = \, 
\frac{2^{3/2}\coupl^{3/2}}{(2\pi)^{3}3^{9/4}}
\left(\frac{\mu}{\Mpl}\right)^{5/2}\frac{\Mpl}{\phihit}\,.
\eeq

After the brane collision, the effective potential consists of 
Eq.~(\ref{eq:potential}) plus the contribution from the newly created strings:
\beq\label{eq:Veff}
V_{\rm eff}(\phi) \, = \, \mu^{10/3}\phi^{2/3}+\coupl n^{(1)}_{\chi} (\phihit-\phi)\,.
\eeq
Let us now check whether the field can come to a rest before it rounds the torus
and hits the $D6$ brane a second time (\ie{} between the values of 
$u_{1,{\rm hit}}$ and $u_{1,{\rm hit}}-1$). A necessary condition for the potential \eqref{eq:Veff}
to develop a local minimum is for $V_{\rm eff}'$ to change sign. As the bare potential \eqref{eq:potential} has a monotonic
$V'>0$, we have to check whether $V_{\rm eff}'<0$ is possible. This amounts to requiring
\beq
\coupl n_{\chi} \, > \, \frac{2}{3}\,\frac{\mu^{10/3}}{\phi^{1/3}}\,,
\eeq
which translates into
\beq\label{eq:stoppingcond}
\frac{\phi}{\Mpl} \, > \, 
\left[\frac{(2\pi)^{3}3^{5/4}}{2^{1/2}\coupl^{5/2}}\right]^{3}
\left(\frac{\mu}{\Mpl}\right)^{5/2}\left(\frac{\phihit}{\Mpl}\right)^{3} \, .
\eeq
For the parameter values of \cite{Eva2} the required field values are
much larger than those in the slow-roll region.
Hence we conclude that the strings produced in a single encounter are too
weak to trap the inflaton field.
 
One may now worry that the buildup of strings produced in various encounters
might trap the inflaton. 
A second encounter will generate new strings between the branes, 
while those from the first collision become heavier and heavier 
\footnote{We ignore the effect of strings reattaching to the 
$D6$ brane at the second collision, which would reduce the attractive 
force between the branes.}. The number density $n_{\chi}$ of strings
is larger at later encounters because the velocity of the inflaton
at the impact point increases [see Eq.~\eqref{eq:phidotphi}].
However, as long as the field remains in the slow-roll region,
the increase in velocity is small. At the same time, however,
space is inflating and thus the number density of the strings is
decreasing exponentially. Thus, strings produced at previous encounters
have a negligible effect on the later encounters as long as the time interval
corresponding to successive encounters is longer than a Hubble expansion
time. The redshifting of the number density of strings also ensures
that the correction to the effective potential has a negligible 
effect on the evolution of the inflaton.
 
We have seen that the strings created after a single collision have a 
negligible effect on the field trajectory. Thus, it is reasonable to 
assume that even after the first brane encounter at $\phihit$ the 
slow-roll trajectory remains valid. For simplicity, let 
us use Eq.~(\ref{eq:phidotphi}) all the way while $\phi \, > \, \phieps$
\footnote{Note that, while Eq.~(\ref{eq:phidotphi}) should be modified after 
each brane collision because of the additional potential terms created by 
each new generation of inter-brane strings, these strings will, if anything, 
\emph{slow down} the brane motion further. Hence, the field in reality would 
be rolling even slower than (\ref{eq:phidotphi}) tells us. Since the string 
production rate $n_{\chi}\propto|\dot{\phi}|^{3/2}$, we are therefore 
\emph{overestimating} the effect of string production.
}.  It is then easy to check for the parameters of \cite{Eva2} 
that no trapping can occur while the inflaton is on the slow-roll trajectory.
Correspondingly, Eq.~(\ref{eq:Nofphi}) tells us how many e-folds are 
produced at each turn. At the beginning of inflation, this number is 
$\Delta N\simeq\mathscr{O}(10)$. It is evident that this expansion of 
space dilutes all the strings created at the first hit to the extent 
that they do not play a role at the second encounter.  However, once 
the expansion drops to $\Delta N<1$ per turn, the created strings may 
indeed accumulate. The crucial question is whether the potential 
contribution (\ref{eq:rhochi}) induced by them can bring the motion of the 
$D4$ to a stop before it reaches the minimum of the original potential 
Eq.~(\ref{eq:potential}) at $\uone=0$ (corresponding to $\phi_{0}=0$). 
If so, 
one may wonder 
how much energy leaks from $\phi$ into the fields localized on the $D6$ 
brane (hence reheating them) while inflation is still under way. We will 
show that even with an overestimation of the string effect its influence is negligible.

\section{Reheat temperature}

In this section, we first estimate the reheat temperature at the final 
brane collision ($\uone=0$) ignoring the effect of string creation at
previous brane encounters, and then refine this calculation taking into 
account the strings created during the last few turns, which do not get 
diluted any more by the inflationary expansion.

\subsection{Neglecting string production}\label{subsec:neglect}

The slow-roll trajectory is valid up to $\phieps$, at which point the kinetic 
and potential energies become equal. Therefore, we can estimate the total 
energy at $\phieps$ as
\beq\label{eq:Etot}
\frac{E_{\rm tot}(\phieps)}{\Mpl^{4}} \, \simeq \, 
2\left(\frac{\mu}{\Mpl}\right)^{10/3}\left(\frac{\phieps}{\Mpl}\right)^{2/3} \, .
\eeq
Since after the breakdown of slow-roll the amount of energy which is lost 
to the expansion of space is negligible [assuming that the reheating
process is rapid, an assumption whose validity is assured by the 
estimate (\ref{eq:comparetH})], the total energy (\ref{eq:Etot})
is approximately conserved down to $\uone=0$. At that point, no inflaton potential 
energy is left [the $D4$ having minimized its worldvolume, see 
Eq.~(\ref{eq:potential})], and hence the entire $E_{\rm tot}$ of 
Eq.~(\ref{eq:Etot}) is converted into kinetic energy. Therefore, the velocity 
of the brane when it reaches $\uone=0$ (corresponding to $\phi_{0}=0$) is given by
\beq
\frac{|\dot{\phi}_{0}|}{\Mpl^{2}} \, = \, 
\sqrt{\frac{2E_{\rm tot}(\phieps)}{\Mpl^{4}}} \, = \, 
2 \left(\frac{\mu}{\Mpl}\right)^{5/3}\left(\frac{\phieps}{\Mpl}\right)^{1/3}.
\eeq

\subsubsection{Reheat temperature from single impact}\label{subsubsec:firstimpact}

To calculate the reheat temperature, let us first determine how much 
energy is channelled from the $\phi$ to the $\chi$ field at the final 
encounter itself. To this end, we set [compare Eq.~(\ref{eq:nchi})]
\beq\label{eq:rhcoupling}
n_{\chi}^{(\uone=0)} \, = \, (\Trh^{\chi})^{3} \, = \, 
\frac{\coupl^{5/2}|\dot{\phi}_{0}|^{3/2}}{(2\pi)^{3}} \, ,
\eeq
assuming that the energy released as $\chi$ particles rapidly
thermalizes (otherwise it would not make sense to talk about a
temperature). Given this assumption, the reheat temperature is
\beq
\Trh^{\chi} \, = \, \frac{\coupl^{5/6}}{2\pi}\,|\dot{\phi}_{0}|^{1/2}
\, = \, \frac{\coupl^{5/6}}{2^{3/4}\pi}\left[E_{\rm tot}(\phieps)\right]^{1/4}.
\eeq
With the estimate (\ref{eq:Etot}), this becomes
\beq\label{eq:Trhchi}
\frac{\Trh^{\chi}}{\Mpl} \, = \, \frac{\coupl^{5/6}}{\sqrt{2}\pi}
\left(\frac{\mu}{\Mpl}\right)^{5/6}\left(\frac{\phieps}{\Mpl}\right)^{1/6}.
\eeq

Note, however, that this only accounts for the energy transferred into the 
$\chi$ field on the first hit after the unwrapping process comes to an end. 
During reheating, the $D4$ oscillates around the $D6$, gradually channelling 
more energy into the $\chi$ field. We can estimate the amplitude of these 
oscillations from (see \cite{Kofman:2004yc})
\beq
\phi_{\rm osc} \, = \, \frac{4\pi^{3}}{\coupl^{5/2}}\,|\dot{\phi}_{0}|^{1/2}\,.
\eeq
We can compare this again to the region in which the particle production 
is effective (compare Sec.~\ref{subsec:fieldtheory}), leading to
\beq
\frac{\Delta\phi}{\phi_{\rm osc}} \, = \, \frac{\coupl^{2}}{4\pi^{3}}\,.
\eeq
For a perturbative value of the coupling, the particle production therefore 
still occurs only during a small fraction of an oscillation.

\subsubsection{Reheat temperature from entire energy transfer}
\label{subsubsec:totaltransfer}

If the $D4$ brane comes eventually to a stop, all of
its energy (apart from its rest mass, which we have consistently ignored in the whole analysis) will go into particles (modulo energy which is lost into
closed string modes, e.g. bulk gravitons \footnote{However, for
the parameters we are using for which the Planck length is smaller than
the string length, the decay into bulk particles will be suppressed.}).
Thus, we can estimate the final reheat temperature by simply
equating the final thermal energy with the inflaton energy at
the beginning of the reheating phase, i.e. using
\beq\label{eq:rhdensity}
\rho_{\rm rh} \, = \, \frac{\pi^{2}}{30}\,g_{*}\,\left(\Trh^{\rm rad}\right)^{4}\,,
\eeq
where $\rho_{\rm rh}$ is the energy density at the beginning of the 
reheating phase. Here, $g_{*}$ is the number of spin degrees
of freedom in the final bath of radiative particles. Taking the
number from Standard Model particle physics, it is a constant of 
$\mathscr{O}(10^{2})$. With Eq.~(\ref{eq:Etot}), this gives a reheat 
temperature of
\beq\label{eq:Trhrad}
\Trh^{\rm rad} \, = \, \left(\frac{60}{g_{*}\pi^{2}}\right)^{1/4}
\left(\frac{\mu}{\Mpl}\right)^{5/6}\left(\frac{\phieps}{\Mpl}\right)^{1/6} \, .
\eeq
We immediately see the same functional dependence on $\mu$ and $\phieps$ as 
in Eq.~(\ref{eq:Trhchi}). Comparing Eq. (\ref{eq:Trhrad}) with this previous 
result, we see that in order to achieve the same total energy transfer, 
the number of oscillations $n_{\rm osc}$ of the $D4$ through the $D6$ can 
be estimated as
\beq\label{eq:nosc}
n_{\rm osc} \, = \, 
\left(\frac{60\pi^{2}}{g_{*}}\right)^{1/4}\frac{\sqrt{2}}{\coupl^{5/6}}\,.
\eeq


\subsection{Cumulative effect of wound strings}
\label{subsec:cumulative}

We now refine the above calculation by including the string-induced corrections. 
Our argument shall be directed towards finding a lower bound on the reheat 
temperature. We will therefore \emph{overestimate} the contribution of the strings that are created during brane collision.
As already argued earlier, as long as each turn corresponds to several e-folds, we can safely assume that all open string states are diluted to a negligible extent. However, as soon as this is not true anymore, some fraction of the strings will survive and get wound around the torus multiple times. (This is the case if we neglect the possibility that they reattach to the $D6$ brane. In principle a 4-6 string can split into a 6-6 string -- that would wind exactly once -- and a massless 6-4 string at the time of the next crossing between the $D4$ and $D6$.) In consistency with overestimating the string effect, we will assume that the point from which on strings do not get diluted completely anymore occurs at some value of $u_1=k$, which is higher than where slow roll breaks down at $u_1=k_{\rm sr}$. Furthermore, we will assume that not only a fraction of them survives, but all of them.

Then our strategy to determine the final velocity $\dot\phi_0$ at $u_1=0$ is the following: We use a slow roll trajectory  all the way down to $u_1=k_{\rm sr}$ from which point on we assume that no energy is lost to the expansion of space anymore. We do, however, include the strings created between the turns $u_1=k$ and the end of slow-roll in the potential 
\begin{eqnarray}
V_{\rm eff}(\phisr)&=&\mu^{10/3}\phisr^{2/3}\label{eq:Vefflast}\\
&+&\,\frac{\coupl^{5/2}}{(2\pi)^3} \sum_{i=\ksr+1}^{k} v^{3/2}(\uone=i) \left[\phi(\uone=i)-\phisr\right] \,. \nonumber
\end{eqnarray}
Note that here we calculate the inflaton velocity $v(\uone=i)$ from the original trajectory \eqref{eq:phidotphi}, which is consistent with our overestimate: a higher velocity corresponds to a higher string production rate. Making use of \eqref{eq:Vefflast} we obtain the inflaton velocity at the end of slow-roll via
\begin{equation}
  \dot\phi_{\rm sr} \,=\, \frac{V_{\rm eff}'}{\sqrt{3V_{\rm eff}}}\,\Mpl\,.
\end{equation}
This enters into the total energy at the point where slow-roll ends
\begin{eqnarray}
&& E_{\rm tot}(u_1=k_{\rm sr}) \,=\, \frac{1}{2}\,\dot\phisr^2 + \mu^{10/3}\phisr^{2/3}\label{eq:Esr}\\
&&\qquad\quad +\,\frac{\coupl^{5/2}}{(2\pi)^3} \sum_{i=\ksr+1}^{k} v^{3/2}(\uone=i) \left[\phi(\uone=i)-\phisr\right] \,, \nonumber
\end{eqnarray}
which we assume to remain conserved from now on. It will mostly be converted into kinetic energy, as the original potential vanishes at $\phi=0$, so the total final energy reads 
\begin{eqnarray}
  E_{\rm tot}(\phi_0=0) &=& \frac{1}{2}\,\dot\phi_0^2 \label{eq:E0} \\
&+&\,\frac{\coupl^{5/2}}{(2\pi)^3} \sum_{i=\ksr+1}^{k} v^{3/2}(\uone=i) \,\phi(\uone=i) \nonumber\\
&+&\,\frac{\coupl^{5/2}}{(2\pi)^3} \,\sum_{i=1}^{\ksr}\, \tilde{v}^{3/2}(\uone=i) \,\phi(\uone=i) \,, \nonumber
\end{eqnarray}
where the last line denotes the contribution from additional strings created during the last few revolutions when slow-roll has ended. These last terms are a bit more complicated to calculate, as we now have to determine the velocity $\tilde{v}$ for $u_1<\ksr$ from energy conservation instead of from the slow-roll trajectory. This is further complicated by the fact that below $\phi_{\rm crit}$ the potential changes to $m^2\phi^2$, see Eq. \eqref{eq:potentialsmall}. However, in the numerical example we study below, it turns out that $\phi_{\rm crit}$ corresponds to less than one revolution in the $u_1$ direction, so we do not have to worry about this fact at all. Also, the iteration necessary to determine the velocities for $\uone<\ksr$ is not as messy as it seems, since in our example $\ksr=2$. 

Equating the energies \eqref{eq:Esr} and \eqref{eq:E0} we finally arrive at the kinetic energy of the inflaton when it reaches the minimum
\begin{eqnarray}
  E_{\rm kin}(u_1=0) &=& \frac{1}{2}\,\dot\phisr^2 + \mu^{10/3}\phisr^{2/3} \label{eq:Ekin0} \nonumber\\
  & - & \frac{\coupl^{5/2}}{(2\pi)^3} \sum_{i=\ksr+1}^{k} v^{3/2}(\uone=i) \,\phisr\\
  & - & \frac{\coupl^{5/2}}{(2\pi)^3} \,\sum_{i=1}^{\ksr}\, \tilde{v}^{3/2}(\uone=i) \,\phi(\uone=i) \nonumber\,.
\end{eqnarray}

With this expression we can now use Eqs.~(\ref{eq:rhcoupling}) or 
(\ref{eq:rhdensity}) to infer the reheat temperature found from our refined 
calculation. Since we have taken into account additional energy loss, the 
result in each case should be a smaller $\Trh$ than those of 
Sec.~\ref{subsec:neglect}. We find
\beqa
\Trh^{\chi}&=&\frac{\coupl^{5/6}}{2^{3/4}\pi}\left[E_{\rm kin}(\phi_0)\right]^{1/4}\quad\textnormal{and}\\
\Trh^{\rm rad}&=&\left[\frac{30g_{*}}{\pi^{2}}E_{\rm kin}(\phi_0)\right]^{1/4},
\eeqa
from the reasoning following a coupling to the $\chi$ field 
(Sec.~\ref{subsubsec:firstimpact}) or energy transfer into 
radiation (Sec.~\ref{subsubsec:totaltransfer}), respectively.

In its full generality, the calculation presented in this subsection seems 
rather involved. However, in a concrete numerical example, the reasoning is 
much more intuitive since the total number $k$ of turns that qualify for 
undiluted string production is small. We now turn to studying such an example 
to illustrate the effect of string production on the reheat temperature in 
the monodromy inflation model at hand.

\section{Numerical example}

We illustrate the relations derived above with a numerical example 
using the parameter values employed in Ref.~\cite{Eva2}. For convenience, 
we summarily list these parameter values along with some useful relations 
in Appendix \ref{app:parameters}. Inserting these values, the critical 
$\uone$ and $\phi$ values from Eqs.~(\ref{eq:ucrit}) and (\ref{eq:defphismall}) 
become
\beq\label{eq:critnum}
\ucrit \, \simeq \, 0.7,\qquad\phicrit \, = \, 0.1\Mpl \, .
\eeq
Therefore, the large field approximation only breaks down during the 
last $\uone$ turn, and we can safely use the corresponding potential 
Eq.~(\ref{eq:potential}) up to that point. From Eq.~(\ref{eq:phieps}) 
it follows that no slow-roll is possible in the region $\phi<\phicrit$ 
with Eq.~(\ref{eq:critnum}).

We set the initial $\uone$ value to be the largest value for which
it makes sense to focus on the dynamics of the candidate inflaton field
alone (see the
corresponding discussion in the first paragraph of Sec.~\ref{sec:inflaton}). 
According to the analysis in \cite{Eva2}, this correspond to a field
value of about $10 \Mpl$. Specifically, we take the first brane
collision to occur at
\beq\label{eq:uin}
u_{1, {\rm hit}} \, = \, 13, \quad\textnormal{corresponding to} \quad
\phihit \, \simeq \, 9.1\Mpl \, .
\eeq
That is, the $D4$ brane is initially wrapped slightly more than 
thirteen times along the 
$\uone$ direction to ensure $\phiin>\phihit$. 

The breakdown of slow-roll occurs [see Eq.~(\ref{eq:phieps})] at the 
field value
\beq\label{eq:srbreakdown}
\phieps \, \simeq \, 0.5\Mpl\,, \qquad \ueps \, \simeq \, 1.8 \, ,
\eeq
and therefore only the last two $\uone$ turns do not occur entirely in 
the slow-roll regime. 

For the scale $\mu$ of the potential, we find from Eq.~(\ref{eq:mu}) that
\beq
\left(\frac{\mu}{\Mpl}\right)^{10/3} \, \simeq \, 8.7\cdot10^{-10}\,.
\eeq
This gives for the initial Hubble scale and velocity
\beq
H_{\rm in} \, \simeq \, 3.6\cdot10^{-5}\Mpl, \qquad 
\dot{\phi}_{\rm in} \, \simeq \, -2.6\cdot10^{-6}\Mpl^2 \, ,
\eeq
illustrating that our non-relativistic treatment of the brane collisions 
is well justified.

Turning to the effective description of the brane collisions, we have to 
specify the coupling between the fields $\coupl$ in addition to the 
parameters of Appendix \ref{app:parameters}. We take
\beq\label{eq:fieldcoupling}
\coupl \, \simeq \, 0.1 \, ,
\eeq
unless otherwise stated\footnote{This does not mean we are equating the string coupling to the coupling that describes the creation of strings on the effective field theory level in general. We just want to assume a reasonable perturbative value for $\coupl$.}. Then, for the first collision occurring at $\phihit$, the interaction takes place 
over a range [see Eq.~\ref{eq:deltaphi2})]
\beq
\frac{\Delta\phi}{\Mpl} \, \simeq \, 5.1\cdot10^{-3} \, ,
\eeq
and from Eq.~(\ref{eq:comparetH}) one obtains that the interaction time is 
$\mathscr{O}(10^{-2})$ smaller than the Hubble time, \ie{} quasi-instantaneous. 

With respect to the question whether the inflaton can get trapped, we keep $\coupl$ 
unfixed for the moment. We argued earlier that Eq. \eqref{eq:stoppingcond} remains valid along the whole slow roll trajectory, which means down to $u_1=2$ in our case.
Inserting our numerical values into 
Eq.~(\ref{eq:stoppingcond}) we find that trapping can occur only if
\beq
\frac{\phi}{\Mpl} \, > \, \frac{0.3}{\coupl^{15/2}} \, .
\eeq
For our choice of the coupling constant $\coupl=0.1$, this corresponds to a
field value much larger than the maximum field value $\phi_{\rm max}$ (see the beginning of Sec.~\ref{sec:inflaton}).
If we were to impose that the field should stop after the first collision of the two branes we would require a coupling between $\phi$ and $\chi$ of $\coupl\approx3$. Even though the string production rate increases towards smaller $\phi$, for the smallest value that is still on the slow roll trajectory (corresponding to $u_1=2$) one would require $\coupl\approx1.4$ for trapping to occur.
Hence for a perturbative field coupling the $D4$ cannot become trapped. One could repeat this estimate for the last two turns, which lie outside the slow roll regime, by assuming that no energy is lost to the expansion of space anymore and that the heavy strings accumulate. However, as we will see shortly, even if we overestimate the string effect its influence remains negligible.

\subsection{Reheating neglecting string production}

Let us first estimate the reheat temperature when the string 
production effects are ignored. Then, considering reheating through 
$\chi$ from Eq.~(\ref{eq:Trhchi}) with values of $\phieps$ and $\mu$ 
corresponding to our parameters, we find
\beq
\Trh^{\chi}\simeq1.6\cdot10^{-4}\Mpl \, \approx \, 3.9\cdot10^{14}{\rm GeV},
\eeq
which is a very high reheat temperature. If we consider the overall 
energy transfer into radiation, the reheat temperature found from 
Eq.(\ref{eq:Trhrad}) is (using $g_{*}=100$)
\beq
\Trh^{\rm rad} \, \simeq2.4 \, \cdot10^{-3}\Mpl\approx5.8\cdot10^{15}{\rm GeV},
\eeq
which would take about [see Eq.~(\ref{eq:nosc})]
\beq
n_{\rm osc} \, \simeq \, 15
\eeq
oscillations around the $D6$ located at $\uone=0$.

\subsection{Reheating with strings from brane collisions}

Clearly, the previous result is an overestimate of the reheat temperature 
since it ignores the energy loss due to string creation and stretching 
between the branes. Another way to see this is as follows: due to the production of strings,
the $D4$ brane will be moving slightly slower towards the end
of the inflationary phase. The slow-rolling approximation will be valid
until a smaller value of the field, and thus at the time of exit
from slow-rolling the $D4$ brane will carry less energy. Let us therefore 
now turn to the more refined calculation of Sec.~\ref{subsec:cumulative} 
in our concrete numerical example.

From Eq.~(\ref{eq:Nofphi}) we can calculate how many e-folds of inflation
are produced at each $\uone$ turn in the slow-roll regime. For example, 
on the first turn between $u_{1,{\rm hit}}=13$ and $\uone=12$, 
$\Delta N_{13\rightarrow12}\simeq 13$ e-folds are produced. However, between 
$\uone=5$ and $\uone=4$, this has dropped to 
$\Delta N_{5\rightarrow4}\approx2$, and in the following turn only one 
e-fold of expansion is produced, $\Delta N_{4\rightarrow3}\approx1$. Hence, 
when looking for a careful reheat temperature estimate, we cannot 
assume that the strings created at and after $\uone=5$ are diluted to a
negligible density. Therefore we set $k=5$ in the notation of 
Sec.~\ref{subsec:cumulative}. We also know, see Eq.~(\ref{eq:srbreakdown}), 
that $\ksr=2$ since $\uone=2$ is the last turn to occur in the slow-roll regime, and 
it corresponds to $\phisr\simeq0.6\Mpl$. 
If we work through the 
numerics, always considering that we are still in the slow-roll regime when 
calculating velocities, we find for the total inflaton energy (\ref{eq:Esr}) that will remain conserved
\beq\label{eq:Etot2num}
E_{\rm tot}(\phisr\simeq0.6\Mpl) \, \simeq \, 7.3\cdot10^{-10}\Mpl^4 \, .
\eeq
There is only one brane collision (at $\uone=1$, corresponding to 
$\phi\simeq0.2\Mpl$) left outside the slow-roll regime
\footnote{Note that we can still use the normalization and potential 
Eqs.~(\ref{eq:defphi}) and (\ref{eq:potential}), respectively, since 
$\ucrit\simeq0.7<1$.}. We need to calculate the velocity $\tilde{v}(u_1=1)$ from energy conservation, i.e. from
\begin{eqnarray}\label{eq:Ekinsr-1}
 E_{\rm kin}(u_1=1)&=& \frac{1}{2}\,\dot\phi(2)^2+\mu^{10/3}\left(\phi(2)^{2/3}-\phi(1)^{2/3}\right)\nonumber\\
 &-&\,\frac{\coupl^{5/2}}{(2\pi)^3}\, \sum_{i=3}^{5}\, v^{3/2}(\uone=i) \left[\phi(2)-\phi(1)\right] \nonumber\\
 &-&\,\frac{\coupl^{5/2}}{(2\pi)^3}\, v^{3/2}(\uone=2) \left[\phi(2)-\phi(1)\right] \,.
\end{eqnarray}
Note that at $u_1=2$ we are just at the end of the slow-roll phase, so we can still obtain $v(u_1=2)$ from the slow-roll trajectory.
From this equation we find for the kinetic energy at the next-to-last brane crossing
\beq
E_{\rm kin}(u_1=1)\simeq8.7\cdot10^{-10}\Mpl^4 \, .
\eeq
This kinetic energy feeds into the calculation of the string production rate at 
the $\uone=1$ encounter 
\begin{equation}
 n_{\chi}(\uone=1)\,=\,\frac{(\coupl \tilde{v}(1))^{3/2}}{(2\pi)^3} \,=\, \frac{(\coupl\sqrt{2 E_{\rm kin}})^{3/2}}{(2\pi)^3}\,. 
\end{equation}
Then, using Eq.~(\ref{eq:Ekin0}) we determine the final kinetic energy at 
$\uone=0$, or $\phi_{0}=0$, respectively, to be
\beq
E_{\rm kin}(\uone=0) \, \simeq \, 7.3\cdot10^{-10}\Mpl^{4} \, .
\eeq
Comparing to Eq.~(\ref{eq:Etot2num}), we see that within our rounding 
accuracy, the entire energy in the system present at $\phi_{\rm sr}$ has 
been converted into kinetic energy at $\phi_{0}=0$. That is, the additional 
potential energy drained by the attached strings is negligibly small. The correction they induce is of the order of $0.2\%$.
Therefore, even after the refined calculation, the high reheat temperature 
estimates found above persist.

This could give rise to a potential gravitino problem in our model. However, the background constructed in \cite{Silverstein:2007ac} breaks supersymmetry at a very high scale, the lowest KK scale. In the case at hand, this corresponds to
\begin{equation}
  m_{\rm KK} \,=\, \frac{2\pi}{\sqrt{\alpha'}\,L_{u_1}}\,,
\end{equation}
as the $u_1$ direction describes the largest extension of the torus [this follows from Eqs. \eqref{eq:beta} and \eqref{eq:Lx}]. With the values from the appendix we obtain $m_{\rm KK}\approx4\cdot 10^{-4}\,\Mpl$, which is about two orders of magnitude smaller than the string scale [see \eqref{eq:ms}]. Since the reheating temperature is slightly larger than the
scale of supersymmetry breaking, there is a potential
gravitino problem (overabundance of gravitinos produced
after reheating \cite{problem}). This is not the topic of our
work. However, we would like to mention that
there are various ways to mitigate the gravitino problem. One
way is to invoke a period of thermal inflation at late times
\cite{thermal}, another one is to make use of nonperturbative
decay channels of the gravitino \cite{nonpert}.

\section{Conclusions}

Recently, it has been proposed to exploit the mechanism of monodromy to 
achieve a large field range for the inflaton in string-motivated models. 
Traditionally, the field range had proven to be generically small in these 
scenarios due to the finite size of the compactified extra dimensions, 
making a sizeable contribution of tensor perturbations (gravity waves) 
to the cosmological perturbation spectrum hard to obtain. Monodromy models 
provide a promising ansatz to overcome this previous phenomenological 
''no-go theorem" for tensor perturbations from string theory.

We have studied the mechanism of reheating in the model proposed in 
Ref.~\cite{Eva2}, modelling the Standard Model by a $D6$ brane at a fixed 
position in the monodromy cycle that the $D4$ brane unwraps while inflation 
is under way. We have shown that there is virtually no energy transferred when 
the branes collide during inflation (even though these collisions occur 
repeatedly), the entire reheating being produced at the last brane encounter. 
This is reassuring in the sense that the additional $D6$ ``stuck in the way'' 
of the inflationary $D4$ does not make it harder to achieve the required number 
of e-folds. We find that the reheat temperature comes out generically 
high in these models. Even for different parameter values one would find that the string production rate is very small due to the small velocity along the slow--roll trajectory. Only if the time between the end of inflation and the last brane crossing becomes considerably longer, then there could be any significant string effects, due to two reasons: first, those strings would be produced at a higher rate because of the larger field velocity and second, they would not be diluted anymore.

We have, of course, studied a rather simplified toy model. It would be interesting to refine our approach by studying a more realistic intersecting brane model, in which the $D4$ would cross a considerably higher number of branes. However, given that we found the open string effect to be negligibly small, we do not expect our conclusions to be altered much if the stacks of branes consist of $N=1,2$, or $3$ $D6$ branes only (these are the numbers needed for the $SU(3)\times SU(2)\times U(1)$ gauge group of the MSSM \cite{Marchesano}). If some of these $D6$ overlap with the $D4$ along one internal direction, we would see another interesting effect emerge -- the $D4$ could actually dissolve into one of the $D6$ (as they are not mutually BPS in this configuration) and form a bound state with considerably lower energy than the initial setup. In this case one would observe the usual tachyonic string modes stretched between non-BPS branes.

Generically one would expect warping in flux compactifications. This has been neglected so far as well as the backreaction of the branes onto the geometry. For a small number of $D4$ and $D6$ this appears to be a valid approach. However, for a large number of $D6$ (they are more massive than the $D4$) it should be taken into consideration.

\enlargethispage{4\baselineskip}
In a variant of the model of Ref.~\cite{Eva2}, it has been proposed 
\cite{McAllister:2008hb} to use an axion field as the inflaton. This axion originates from the NS or RR two-form, which is integrated over a two-cycle in the internal geometry and therefore appears as a scalar in the four-dimensional theory. The usual shift symmetry of these fields is broken in the presence of branes, which makes them axionic inflaton candidates. Reheating would then proceed via the closed string sector and requires a different description than the simple field theoretic ansatz we used.

\begin{acknowledgments}

We wish to thank Keshav Dasgupta and Andrew Frey for useful discussions.
The work is supported in part by NSERC Discovery Grants and by the Canada 
Research Chairs program. LL acknowledges support through a PhD scholarship of the German Merit Foundation.

\end{acknowledgments}
\begin{appendix}
\section{Model parameters}\label{app:parameters}

In this appendix, we cite important relations between background parameters 
along with their default values (taken from Ref.~\cite{Eva2}), which we use 
whenever numerical estimates are carried out.

\subsection{Useful relations}

The string length and the 4d Planck mass are related by
\beq\label{eq:alphap}
\frac{1}{\alphap} \, = \, \frac{(2\pi)^{7}\gs^{2}}{L^{6}}\,\Mpl^{2}\,,
\eeq
where the radial modulus $L$ is a measure of the volume of the torus 
we are considering (whose coordinates are $(x,u_1,u_2)$). 
We have
\beq
L^{3} \, = \, L_{u}^{2}L_{x}\,,
\eeq
and since we have two copies of this torus (with an orbifold projection), the total compact volume is $V=L^6/2$.
$L_{x}$ is the length scale in the $x$ direction [see Eq.~\eqref{metric}], and $L_u$ is an averaged length scale in the $\uone,u_{2}$ directions
given by
\beq
L^{2}_{u} \, = \, L_{\uone}L_{u_{2}}\,.
\eeq
We define the anisotropy parameter $\beta$ via
\beq\label{eq:beta}
\beta \, = \, \frac{L_{u_{2}}}{L_{\uone}} \, = \, 
\frac{L_{u}^2}{L_{\uone}^{2}} \, = \, \frac{L_{u_{2}}^{2}}{L_{u}^{2}}\,.
\eeq

\subsection{Background parameter values}
\label{subsec:bgvalues}

The length scales can be expressed in terms of the background flux quantum numbers $M$ and $K$ as (see \cite{Silverstein:2007ac})
\begin{eqnarray}
L \, &=& \, c_{L}\cdot K^{1/6},\label{eq:L}\\
L_{x} \, &=& \, c_{L_{x}}\cdot M^{-1/2},\label{eq:Lx}\\
L_{u} \, &=& \, \frac{c_{L}^{3/2}}{c_{L_{x}}^{1/2}}\,\left(KM\right)^{1/4}.
\label{eq:Lu}
\end{eqnarray}
The coefficients $c_{L}$ etc. were chosen with numerical values
\beq
c_{L} = 1.7,\quad c_{L_{x}} = 8.6,\quad 
\left(c_{L}^{3/2}/c_{L_{x}}^{1/2}\right)\simeq 0.75 \, .
\eeq

Let us now list the values of $\beta$ and of the fluxes $M,K$ which
were used in \cite{Eva2} to obtain an inflationary model with a
sufficient number of e-foldings of slow-roll inflation and with
a correct normalization of the power spectrum of cosmological
perturbations, values which we use in the text for our numerical estimates:
\begin{eqnarray}
\beta \, &\simeq& \, 0.04\\
M \, &\simeq& \, 1\\
K \, &\simeq& \, 2.2\cdot10^{6}\,.
\end{eqnarray}
From (\ref{eq:L})-(\ref{eq:Lu}) this leads to the scales
\beq
L\simeq19.4,\quad L_{x}\simeq8.6,\quad L_{u}\simeq 29.1\,.
\eeq
Note that this implies in particular that $L_{u_1}\simeq 145.5$ is the greatest length scale, because $\beta$ is small.

The string coupling amounts to
\beq
  \gs \, \simeq \, 0.1\,,
\eeq
and for the ratio between the string and the Planck scale we find [see Eq.~(\ref{eq:alphap})]
\beq\label{eq:ms}
\sqrt{\alphap}\,\Mpl \,=\, \frac{\Mpl}{M_s} \, \simeq 117\,.
\eeq

\end{appendix}


\end{document}